\documentclass[reprint,amsmath,amssymb,aps,prxq,nofootinbib]{revtex4-2}
\usepackage{graphicx}
\usepackage{hyperref}

\begin{document}
	
\title{Increased-Efficiency Multiple-Decoding-Attempts Error Correction \\ for Continuous-Variable Quantum Key Distribution}

\author{Lukas Eisemann}
\email{lukas.eisemann@fau.de}
\author{Ömer Bayraktar}
\author{Stefan Richter}
\author{Kevin Jaksch}
\author{Hüseyin Vural}
\author{Christoph Marquardt}
\affiliation{Chair of Optical Quantum Technologies, Friedrich-Alexander-Universität Erlangen-Nürnberg, Staudtstr.~7 / A3, 91058 Erlangen, Germany}
\affiliation{Max Planck Institute for the Science of Light, Staudtstr.~2, 91058 Erlangen, Germany}

\begin{abstract}
In continuous-variable quantum key distribution (CV-QKD), the performance of the information reconciliation (IR) step is critical for the achievable secret key rate (SKR) and transmission distance. 
We show how to improve on the recently introduced implementation of an IR-protocol involving multiple decoding attempts (MDA) and validate the method on simulated data in different application scenarios.
Throughout, we demonstrate meaningful SKR-gains compared to both the standard protocol of a single decoding attempt and to the original MDA-implementation, even at given decoding complexity.
\end{abstract}

\maketitle

\section{Introduction}

Quantum key distribution (QKD) provides 
an information-theoretically secure
way of growing keys 
shared by two legitimate parties.
In typical prepare-and-measure implementations, this entails Alice (A) transmitting a random sequence of non-orthogonal quantum states to Bob (B).	
After this transmission, A and B hold correlated but unequal bit strings. This necessitates the 
use of error correction (EC, or information reconciliation, IR) protocols, which attempt to
correct frames
at the price of revealing part of the information of the frames, i.e., at the price of a classical leakage. 
Typically, the performance of the EC-step is strongly affecting the SKR and secure transmission distance, and the design of EC codes (ECCs) and protocols is subject of active research. In particular, due to the typical low signal-to-noise-ratios (SNRs), this is the case for continuous-variable (CV) QKD, which promises to allow for realizations of QKD on existing telecom infrastructure\cite{reviewQKD, usenkoReviewCV, reviewCVEC}.

Recently, an ECC in the class of 
so-called raptor-like low-density parity-check (RL-LDPC) codes
has been developed with a range of code-rates $r$ suitable for 
CV-QKD \cite{rlBeijing}. 
Such codes come with an integrated method of rate-adaptation and allow for good performance across a wide range of $r$.
In  \cite{ecc}, a RL-LDPC code with the to date largest $r$-range (1$\%$ to 20$\%$) has been designed, and an implementation of it has been made available 
open-access \cite{ecc,toolkit}, removing the barrier of code-design.

Rate-adaptability allows to improve the performance of the EC step in two ways. One is by tuning the code-rate to its optimal value \cite{reviewCVEC}, which is best done by jointly tuning the SNR \cite{jointOptimiz}. 
Secondly, however, in case of a decoding failure, it can allow for further decoding attempts (DAs) on that frame if the method of rate-adaptation avoids unnecessary classical leakage in the process.
This second way
has been pioneered in \cite{mda}, where the approach was dubbed Multiple-Decoding-Attempts (MDA) protocol. The method of 
rate-lowering 
employed in \cite{mda} is bit-revelation \cite{spProtocolDV,spProtocolCV}, and significant SKR gains compared to the standard protocol of a single decoding attempt (SDA) have been demonstrated.

In this work, we expand on the formalization of MDA-protocols and, based on that, point out how to improve on the 
original implementation of \cite{mda}. To demonstrate this quantitatively, we choose an MDA-implementation based on 
the rate-adaptation inherent to 
RL-LDPC codes.
For the validation on simulated data, we consider a use-case of MDA that is analogous compared to the one presented in \cite{mda} as well as a qualitatively new use case.
We also expand on the analysis of decoding complexity by resolving results both in terms of a heuristic measure as used in \cite{mda} and directly in terms of the average number of decoding iterations.

The remainder of this paper is organized as follows.
General theoretical considerations on MDA-protocols are provided in \ref{ssec:general} and the definition of an implementation based on RL-LDPC codes is presented in \ref{ssec:implementations}.
Sec.~\ref{sec:simulated} presents the application to simulated data, with the two use-cases treated in \ref{ssec:2examples} and the resolution of performance in terms of decoding complexity presented in \ref{ssec:complexity}.
Our findings are summarized in Sec.~\ref{sec:conclusion}.

\section{Multiple-Decoding-Attempts Protocols}\label{sec:mda}
\subsection{General Aspects}\label{ssec:general}
	
Following \cite{mda}, we assume that the asymptotic fraction of distilled key bits per sent symbol in the standard protocol of a single decoding attempt (DA) is given by\footnote{The global factor $(1-\text{FER})$ in Eq.~\eqref{skf} is appropriate for protocols in which all frames that have not been successfully reconciled are discarded (see, e.g., \cite{Jain, ferPirandola}).}\footnote{In \cite{johnson}, Eq.~\eqref{skf} has been argued to be inconsistent in the regime of $\beta>1$ and high modulation variance $V_A$ for the case of Gaussian modulation.}
	\begin{equation}\label{skf}
		K
		= 
		\left( 1 - \text{FER} \right) \left( \beta I_{\text{AB}} - \chi \right) \, .
	\end{equation}
Here, $\text{FER}$ is the frame error rate, i.e. the fraction of frames not successfully corrected; 
$I_{\textit{AB}}$ is the mutual information between A and B and defines the information-theoretic lower bound on the classical leakage for successful decoding;
$\chi$ is the Holevo-information, which quantifies the 
quantum leakage, i.e., the upper bound on the information on the raw key gained by Eve during her attack on the quantum channel.
The actual classical leakage achieved is quantified by $\beta$. 
Assuming heterodyne detection and a SNR too low for encoding more than 2 raw key bits per symbol\cite{sec}, one has $\beta= 2 r /I_{\textit{AB}}$, where $r$ is the code rate of the (binary) ECC used. The leakage during the error correction protocol is then given in 
terms of $r$ 
as $L_{cl}=2(1-r)$.
		
Now let us consider an MDA-protocol, i.e., a protocol that allows further DAs on the frames that have not been corrected successfully during the first DA.
The obstacle to overcome for
any such protocol is the potential accumulation of classical leakage.
Let us consider the example of a second DA on a set of frames not successfully corrected during the first DA.
Consider selecting for the two DAs two codes from a set of fixed-rate binary LDPC codes with rates $r_1 > r_2$.
Eve obtains information about the frames during both DAs 
via the parity-check (PC) results publicly communicated in order to facilitate the EC. 
While as stated this classical leakage amounts to  $2(1-r_1)$ after the first DA,
it may grow beyond $2(1-r_2)$ after the second DA to the extent the parity checks (PCs) of the two codes used are linearly independent.
The extra classical leakage due to a second DA thus only saturates
	\begin{equation}\label{extraLeakage}
		\Delta L_{cl} \geq 2(r_1 - r_2) \,,
	\end{equation}
if the two codes are maximally related. 
Such minimization of the extra classical leakage in lowering the code rate between DAs can be taken as the definition of \textit{true} rate adaptability as opposed to effective rate adaptability provided by a dense set of generic fixed-rate codes as considered above.

For a MDA-implementation 
based on true adaptation of $r$,
the asymptotic secret fraction after $k$ DAs can be improved to \cite{mda}
	\begin{align}\label{skfMDA}
		K_k
		= 
		&\left( 1 - \text{FER}_1 \right) \left( \beta_1 I_{\textit{AB}} - \chi \right) +
		\nonumber
		\\
		+ &\sum_{i=2}^k \prod_{j=1}^{i-1} \text{FER}_j \left( 1 - \text{FER}_i \right) \left( \beta_i I_{\textit{AB}} - \chi \right) \, ,
	\end{align}
where the $\text{FER}_j$ with $j\geq2$ are defined only on the set of frames that are not successfully corrected during the previous DA $j-1$. Thus the overall frame error rate after the $j$th DA is given by $\prod_{m=1}^{j} \text{FER}_m$.
In terms of the secret key fraction (SKF) and FER achieved after $k$ DAs, one may also define an effective value of the efficiency, $\beta_{\textit{eff}}$, by
\begin{equation}\label{betaEff}
	K_k \equiv \left( 1 - \prod_{m=1}^{k} \text{FER}_m \right) \left( \beta_{\textit{eff}} I_{\textit{AB}} - \chi \right) \, ,
\end{equation}
as an alternative measure of the improvement due to MDA.
	
From Eq.~\eqref{skfMDA} it can be seen that the potential relative gain is certainly limited as 
	\begin{equation}\label{boundGain}
		\frac{K_k}{K_1} -1 < \frac{\text{FER}_1}{1- \text{FER}_1} \,.
	\end{equation}
One way in which $\text{FER}_1$ can take a high value 
(even when $r_1$ can be freely chosen) 	
is in the presence of fluctuating channel parameters.
Such fluctuations cannot be avoided, e.g., in CV-QKD over atmospheric channels \cite{ruppert}.
In such a scenario, MDA can act as an insurance.
But a high value of $\text{FER}_1$ may also be intentional:
\begin{itemize}
	\item[\textit{i})] Parameters may be such that the optimal frame error rate $\text{FER}_*$, i.e., the FER associated with $r_*$, the choice of $r$ that maximizes the SKF after one DA, is high. In that case, one may choose $r_1 \geq r_*$ to achieve $\text{FER}_1 \geq \text{FER}_*$. 
	\item[\textit{ii})] When $\text{FER}_*$ is low, one may still choose a value of $r_1$ that results in a high $\text{FER}_1$. 
\end{itemize}
We are going to demonstrate the application of MDA in scenarios \textit{i}) and \textit{ii}) in \ref{sec:simulated}.

\subsection{Implementations}\label{ssec:implementations}
As pointed out in the previous section, any method of true rate adaptation (as defined around Eq.\eqref{extraLeakage}) is suitable for an implementation of MDA.
The difference in performance of two MDA implementations, assuming the \textit{same} ECC of $r_1$ used in the first DA, therefore depends on the respective FER-performance under rate-lowering.

In the original implementation of \cite{mda}, 
minimal extra leakage $\Delta L_{cl}$ is achieved by lowering the code rate via bit-revelation \cite{spProtocolDV,spProtocolCV}.
The bits to be revealed are selected randomly and their index along with their value are publicly communicated.
In reverse reconciliation, Alice can then utilize this information by setting the corresponding log-likelihood-ratios (LLRs) to large values, thereby aiding the decoding of the entire frame \cite{mda}.

The alternative method of rate-adaptation we consider in the present paper is that inherent to RL-LDPC codes, which likewise avoids unnecessary classical leakage during rate-lowering, but also promises better performance as these codes are optimized for the entire range of code rates.
The barrier of designing a RL-LDPC code is removed by the open-access availability \cite{ecc,toolkit} of the code presented in \cite{ecc}. 
	
The code-rate of a RL-LDPC code can be chosen by using only appropriate portions of a single parity-check matrix (PCM) $H_{t}$.
Starting from a certain portion $H$ in the upper left corner of $H_t$, the rate $r$ can be lowered by ''uncovering`` an equal number $d$ of rows and columns of $H_t$ \cite{rlBeijing}.
I.e., with $m$ and $n$ the number of rows and columns of $H$, the adapted PCM $H^\prime$ consists of $m^\prime = m+d$ rows and $n^\prime = n+d$ columns, and the	
resulting lower rate is given by $r^\prime =  (n^\prime-m^\prime)/n^\prime = (n-m)/(n+d)$. This is in contrast to the technique of bit-revelation, where 
the frame length remains unchanged, $n^\prime = n$,  and thus
for a number of $d$ revealed bits one has $r^\prime = (n-m-d)/n$. 
Therefore, the two implementations require different values of $d$
in order to realize the same decrease in code rate $r-r^\prime$.

The minimization of extra classical leakage $\Delta L_{cl}$ in the case of MDA based on RL-LDPC codes can be seen clearly when considering the extreme case of $d=1$.
The design of $H_t$ is such that any additional column is nonzero only in the last entry, $H^\prime_{i,n+1} = \delta_{i,m+1}$.
Thus, the newly added bit enters only the newly added parity check (PC). In other words, among the $m+1$ PCs, only the last one contains new information, and the information contained in the preceding $m$ PCs is identical with the information about the first $n$ bits already leaked during the first DA.
By extension to an arbitrary value of $d$, one sees that extra leakage is indeed minimized.

One may argue that the attribute ''raptor-like`` in the name ''RL-LDPC`` already suggests the application of such codes within a MDA scheme since, effectively, raptor-codes likewise allow to keep lowering the code-rate until successful decoding (and have been applied to CV-QKD \cite{raptor}). In fact, the authors of \cite{mda} have 
pointed out this analogy between the MDA-approach and a reconciliation based on raptor-codes.
	
In either of the above two MDA-schemes, communication between A and B is necessary between DAs. For reverse (forward) reconciliation, A (B) needs to inform B (A) about decoding failures. B (A) then provides information to A (B) to facilitate the next DA at a lower code rate: 
for MDA based on bit revelation, this is done by the already mentioned communication of positions and values of the bits to be revealed;
for MDA based on RL-LDPC codes (or, more precisely, based on the method of rate-adaptation inherent to those ECCs), the extension of the syndrome has to be communicated. 
As has been argued in \cite{mda}, compared to the standard SDA protocol, there is no significant additional latency created as a result of this.

\section{Application to Simulated Data}\label{sec:simulated}
	
In this section, we quantitatively investigate the performance of the proposed MDA-variant based on the simulated transmission of frames.
In exemplary applications, we quantify the gains in SKF both compared to a single decoding attempt (SDA) and compared to revelation-based MDA (Sec.~\ref{ssec:2examples}).
We extend the analysis by additionally considering decoding iterations in Sec.~\ref{ssec:complexity}.
For brevity, from now on we are going to refer to the MDA-implementation based on bit-revelation and that based on the rate-adaptation inherent to RL-LDPC codes as $\text{MDA}_a$ and $\text{MDA}_b$, respectively.

For a fair comparison, we use the \textit{same} ECC (the RL-LDPC code from \cite{ecc}) for SDA as well as for \textit{both} MDA implementations, the difference being only in the method of rate-adaptation used for extra decoding attempts.

\subsection{Two scenarios}\label{ssec:2examples}
	
\begin{figure}
\centering
\includegraphics[width=0.49\textwidth]{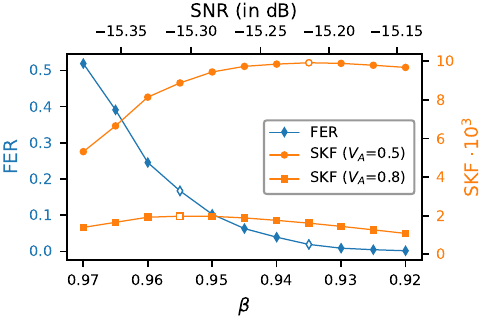}
\caption{Single decoding attempt: Frame error rate (FER) and asymptotic secret key fraction (SKF) (with the RL-LDPC code of \cite{ecc} with the fixed choice of $r=0.02$) on frames simulated for a range of SNR-values. 
The asymptotic SKF is shown for the two choices of the modulation variance $V_A$ considered, resulting in \textit{i}) a high and \textit{ii}) a low value of $\text{FER}_*$ (white color markers), respectively.}
\label{fig:range}		
\end{figure}

We are going to consider two different application scenarios, in which $\text{FER}_*$ (i.e., the optimal value of the FER in the SDA protocol) is \textit{i}) high and \textit{ii}) low (corresponding to the cases outlined in Sec.~\ref{sec:mda}).

For either case, we are
going to consider the simplest MDA-variant, namely $k=2$, i.e., in case of a decoding failure, we admit only one additional DA.
	
We 
consider the case of QPSK-modulation, heterodyne detection, and reverse reconciliation with usage of the full soft information as side-information for decoding. Details are presented in the appendix.

In order to obtain the performance of the SDA protocol as a reference result, we choose as code rate $r=0.02$ 
and (allowing $l_{\textit{max}}=500$ decoding iterations) find the $\text{FER}$ and asymptotic SKF in a Monte-Carlo simulation for a set of $\text{SNR}$-values, see Fig.~\ref{fig:range}. 
Following \cite{mda}, the calculation of the SKF is based on the security proof of \cite{zhangHolevo},\footnote{Security proofs for DM-CV-QKD are subject of active research and more advanced proofs have been developed\cite{usenkoReviewCV} (for implementations of the to-date most general finite-size proof \cite{fsProof}, see \cite{Jaksch2024,Hajomer2025,16qam25km,Ng2025}). The choice of proof made in \cite{mda} may be considered pragmatic for the purpose of demonstrating the gains due to MDA, and we follow this choice for the sake of comparability.} with the SNR varied solely due to variation of the transmission loss.
The frame length (which is required by the RL-LDPC code for the chosen value of the code-rate $r$) is $n=10^6$. Further details are provided in the appendix.

The appropriate mutual information $I_{\textit{AB}}$ for the virtual channel in the reconciliation employed is that of the BI-AWGN-channel. At the $\text{SNR}$-values under consideration, this is however well approximated by the capacity of the Gaussian channel (for heterodyne detection, i.e., per symbol),
	\begin{equation}\label{cap}
		I_{\textit{AB}} \approx 
		\log_2\left( 1 + \text{SNR} \right) \,,
	\end{equation}
and we shall use this approximation, following \cite{mda}.

In Fig.~\ref{fig:range}, the SKF is plotted for two different choices of the modulation variance $V_A$. Each choice results in a different $\text{SNR}_*$, i.e., a different SNR-value for which the asymptotic SKF \eqref{skf} takes its maximum value $K_*$ (within the granularity used). The associated value $\text{FER}_*$ in the two cases is once relatively low and once relatively high. 
These are the two exemplary scenarios in which we are going to apply both 2-DA schemes (in \ref{sssec:ex1} and \ref{sssec:ex2} below). The asymptotic SKF after two DAs according to Eq.~\eqref{skfMDA} is given by
\begin{align}\label{skfGain}
\nonumber
K_2^{(a/b)} = &\left( 1 - \text{FER}_1 \right)\left(2 r_1 -\chi \right) +
\\
+ &\text{FER}_1 \left( 1 - \text{FER}_2^{(a/b)} \right) \left(2 r_2 - \chi  \right) \,,
\end{align}
where the superscript $a$ or $b$ indicates the 2-DA scheme, i.e., the method of rate adaptation used to lower the code rate from $r_1$ to $r_2$.

In general, the gain $K_2 - K_*$ 
depends on the choices for $r_1$ and $r_2$, or since $\text{SNR}_*(V_A)$ is fixed, equivalently, on $\beta_1$ and $\beta_2$. 
Regarding the optimal choice of $r_1$ and $r_2$, qualitatively, it can be expected that $r_1-r$ is positive and the larger, the smaller is $\text{FER}_*$, while $r_2-r$ is negative and the larger (in absolute value), the larger is $\text{FER}_*$.
In the following, we are going to determine the optimal choices numerically.
For each of the two DAs, we allow $l_{\textit{max}}=400$ decoding iterations.
In \ref{ssec:complexity}, we also vary $l_{\textit{max}}$ in order to compare the complexity of the schemes used.

For a motivation of the specific values of $V_A$ considered here, see the appendix.

\subsubsection{High $\text{FER}_*$}
\label{sssec:ex1}

For the case of $V_A=0.8$, we choose $r_1=r\equiv0.02$ or, equivalently, $\beta_1 = \beta_* = 0.955$, following \cite{mda}. As a benefit, this leaves only $r_2$ to optimize, which is conveniently parametrized as $2 r_2 \equiv (\beta_1- \Delta\beta)I_{\textit{AB}}$.

	\begin{figure}[]
	\centering
	\includegraphics[width=0.49\textwidth]{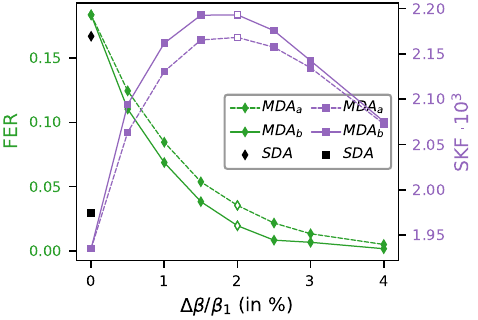}	
	\caption{Example \textit{i}): high $\text{FER}_*$ ($V_A = 0.8$): Frame error rate and secret fraction after the second decoding attempt (DA) depending on the step in code-rate. 
    The values plotted at $\Delta \beta =0$ show the result after the first DA and the best possible SDA result, respectively.} 		
	\label{fig:plotEx1}
	\end{figure}
	\begin{figure}[]
		\centering
		\includegraphics[width=0.49\textwidth]{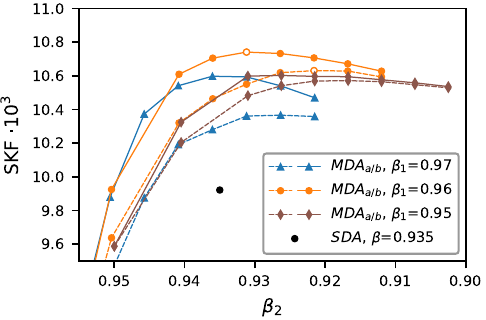}		
		\caption{Example \textit{ii}): $\text{FER}_* \ll 1$ ($V_A=0.5$): Secret key fraction (SKF) after the second decoding attempt (DA) depending on the choices of code rate for either of the DAs. The best possible result of a single decoding attempt (SDA) is shown as a reference.}
		\label{fig:plotEx2}
	\end{figure}

In Fig.~\ref{fig:plotEx1}, we show the secret fraction and the frame error rate after the second DA, $K_2$ and $\text{FER}_1 \, \text{FER}_2$, respectively, 
as a function of the code-rate step, quantified in terms of $\Delta\beta /\beta_1 = 1-\beta_2/\beta_1$, which in the case of $\text{MDA}_a$ corresponds to the fraction of information bits revealed.
The values plotted at $\Delta \beta =0$ show the result after the first DA, $K_1$ and $\text{FER}_1$, which despite to the choice $\beta_1 = \beta_*$ are slightly worse than the SDA result due to the difference in $l_{\textit{max}}$.

The $\text{FER}$-improvement is of course monotonically growing with $\Delta \beta$, but the information advantage suffers from the $\Delta \beta$-increases.

We see that for either 2-DA scheme, $K_2^{(a/b)}$ is the largest for the choice $\Delta \beta /\beta_1 = 2 \%$.
With that choice of $\beta_2$, the 2-DA protocols manage to lower the overall FER from $16.7 \%$ to $2.0 \%$ ($3.5 \%$) at the price of lowering the effective efficiency (cf. Eq.~\eqref{betaEff}) from $95.5\%$ to only $\beta_{\textit{eff}} \approx 95.2$ in either case.

In terms of the key rate itself, the gain of $\text{MDA}_b$ compared to the SDA-protocol is given by
    \begin{equation}\label{gB}
	g_b \equiv \frac{K_2^{(b)}}{K_*} - 1 \approx 11.1 \% \,.
	\end{equation}
By comparison, 
$\text{MDA}_a$ yields
\begin{equation}\label{gA}
	g_a \equiv \frac{K_2^{(a)}}{K_*} - 1 \approx 9.8\% \,.
\end{equation}
Thus, changing from scheme $a$ to $b$ further improves the SKF 
by $g_b-g_a \approx 1.3\%$ (of $K_*$)
or, equivalently, improves the 2-DA gain by
    \begin{equation}\label{iABex1}
	\frac{g_b}{g_a} - 1 \approx 12.9 \% \,,
	\end{equation}
which constitutes a first exemplary quantification of the gains made available by the two MDA-implementations under comparison.

As a side remark, the gain \eqref{gA} is higher than that found in \cite{mda} despite the value of $\text{FER}_*$ here being slightly smaller. This can be explained by the better performance of the ECC used here, in particular, by its lower error floor, and has been expected by the authors of \cite{mda}.

\subsubsection{Low $\text{FER}_*$}    
\label{sssec:ex2}

For the case of $V_A=0.5$, as can be seen from Fig.~\ref{fig:range}, $\beta_*=0.935$. In contrast to the previous example, the choice $\beta_1 = \beta_*$ (i.e., $r_1=r\equiv 0.02$), which leads to $\text{FER}_1\approx\text{FER}_*$, is not a viable option since now $\text{FER}_*\approx 2 \%$, leaving little room for improvement.
For this scenario, we independently vary $r_1$ and $r_2$ to numerically determine the optimal choice, where as before we vary $\Delta \beta/\beta_1$ in steps of $0.5 \%$.
The result is summarized in Fig.~\ref{fig:plotEx2} (where now the FER-improvements are omitted). 
As can be seen, the preferred parameter choices are $\beta_1=0.96$ and $\Delta \beta/\beta_1 = 3\%$ ($4\%$) for $\text{MDA}_b$ ($\text{MDA}_a$).
With those choices, the 2-DA protocols manage to increase the effective efficiency (cf. Eq.~\eqref{betaEff}) from $93.5 \%$ to $\beta_{\textit{eff}}\approx 95.2 \%$ ($95.0 \%$) while slightly decreasing the already low FER from $2.0\%$ to $1.1\%$ ($1.2\%$)

The key rate gains defined in Eq.s \eqref{gB} and \eqref{gA} in this scenario amount to
\begin{equation}
    g_b=8.3\% \,, \quad g_a=7.1\% \,, 
\end{equation}
implying an improvement among the 2-DA schemes of $g_b-g_a\approx 1.2\%$ or, in relative terms,
\begin{equation}\label{iABex2}
\frac{g_b}{g_a} - 1 \approx 15.5 \% \,.
\end{equation}

\subsection{Complexity}\label{ssec:complexity}

\begin{figure*}
	\begin{minipage}{\columnwidth}
		\centering
		\includegraphics[width=\textwidth]{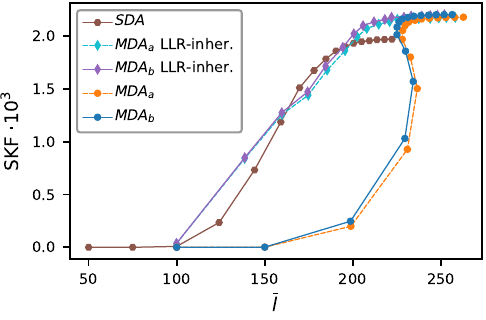}
	\end{minipage}
	\hfill
	\begin{minipage}{\columnwidth}
		\centering
		\includegraphics[width=\textwidth]{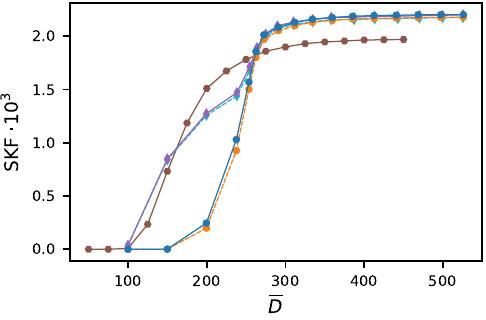}
	\end{minipage}
	\caption{Performance and complexity: SKF achieved by the various schemes versus the average number of decoding iterations used, $\overline{l}$, and the average allowed number of decoding iterations, $\overline{D}$ (see Eq.~\eqref{DBar}), respectively. The plotted points result from a variation of $l_{\textit{max}}$ from $50$ to $450$ in steps of $25$.}
	\label{fig:complexitySKF}
\end{figure*}

\begin{figure}[]
	\centering
	\includegraphics[width=0.49\textwidth]{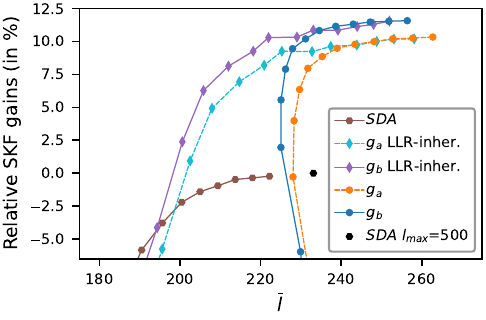}
	\caption{Zoom into high-$\overline{l}$ range of Fig.~\ref{fig:complexitySKF} with the SKFs 
    re-expressed in terms of the gains relative to the SDA result for $l_{\textit{max}}=500$ from Fig.s \ref{fig:range} and \ref{fig:plotEx1}.
    }	
	\label{fig:complexityUsedSKFZoom}
\end{figure}

For a fair comparison of MDA to SDA as well as of the two MDA implementations to each other, not only the performance but also the decoding complexity should be considered.
For that, we revisit the first example (Fig.~\ref{fig:plotEx1}), i.e., the case of $V_A=0.8$ with the choices $\beta_1=\beta_*=95.5 \%$ and $\Delta \beta /\beta_1 = 2 \%$.

To assess the complexity of decoding, the most relevant quantity is arguably the average number of decoding iterations used, $\overline{l}$.
In addition, following \cite{mda}, one may define the heuristic quantity
\begin{equation}\label{DBar}
\overline{D} \equiv  
l^{(1)}_{\textit{max}} + 
\sum_{i=2}^k l^{(i)}_{\textit{max}} \prod_{j=1}^{i-1} \text{FER}_j \,,
\end{equation}
which may be dubbed the average allowed number of decoding iterations.
For the SDA protocol, $\overline{D}=l_{\textit{max}}$, while for a 2-DA protocol with the choice $l^{(1)}_{\textit{max}}=l^{(2)}_{\textit{max}}=l_{\textit{max}}$, one has $\overline{D}=l_{\textit{max}} \left( 1 + \text{FER}_1  \right)$. 

For the simulations in Fig.s \ref{fig:range} and \ref{fig:plotEx1}, as stated above, we have allowed $l_{\textit{max}}=500$ for the SDA scheme and $l_{\textit{max}}=400$ for each DA in the 2-DA schemes, implying $\overline{D} \approx 473$ in the latter case.
On the other hand, the average number of decoding iterations used in those simulations is $\overline{l} \approx 233$ for the SDA scheme as well as $\overline{l}_b \approx 247$ and $\overline{l}_a \approx 253$ for the two 2-DA schemes respectively.

For a more systematic comparison in the following, we vary the value of $l_{\textit{max}}$ from $50$ to $450$ in steps of $25$ (maintaining $l^{(1)}_{\textit{max}}=l^{(2)}_{\textit{max}}=l_{\textit{max}}$). For those values, we plot the achieved SKFs as a function of both $\overline{l}$ and $\overline{D}$ in 
Fig.~\ref{fig:complexitySKF}. 
Analogous plots of the FERs (Fig.~\ref{fig:complexityFER}) 
can be found in the appendix.

For comparison, following \cite{mda}, for both 2-DA schemes, we also implement inheritance of the log-likelihood-ratios (LLRs) between the DAs, i.e., in the case of an unsuccessful first DA, the second DA does not start with the channel-LLRs but instead with the LLR-values of the first DA after iteration $l_{\textit{max}}$ 
(see \cite{mda} for details).

For better visibility, in Fig.~\ref{fig:complexityUsedSKFZoom}, we show a zoom into higher $\overline{l}$-values of Fig.~\ref{fig:complexitySKF} in terms of the SKF gains relative to the best possible SDA result for $l_{\textit{max}}=500$ (cf. Eq.s \eqref{gB} and \eqref{gA}).
As can be seen from Fig.s \ref{fig:complexitySKF} and \ref{fig:complexityUsedSKFZoom}, without LLR-inheritance, the better performance of the two 2-DA protocols as compared to the SDA-protocol comes at a slight price of decoding complexity. This is however no longer the case when LLR-inheritance is employed. 
Comparing the 2-DA protocols with each other, $\text{MDA}_b$ can be observed to maintain its edge over $\text{MDA}_a$ 
at both given $\overline{l}$ and given $\overline{D}$.

In addition, following \cite{mda}, we implement an early termination (ET) algorithm of the decoder: If the output bits for the decoded frame do not vary in five successive decoding iterations, the DA is terminated, assuming a decoding failure.
Remarkably, in contrast to the results of \cite{mda}, the frequency of ET we observe is extremely low and yields no significant gains. This appears to indicate that the structure of the ECC \cite{ecc} is such that Sum-Product-Algorithm-decoding rarely gets stuck in the above sense.

\section{Summary and Conclusions}\label{sec:conclusion}

The achievable key rate in CV-QKD strongly depends on the performance of the IR step. 
A whole chain of potential improvements of the IR performance is tied to 
adaptation
of the code rate.
Within the ubiquitous protocol of a single decoding attempt (SDA), rate-adaptability in principle allows for realizing a value of the code rate close to the optimal one. Since the latter depends on the SNR, this is especially important in the face of fluctuating channel parameters.
Rate-adaptability also allows for protocols using multiple decoding attempts (MDA), which have been introduced and demonstrated superior to the SDA protocol in \cite{mda}.
The performance of a given MDA implementation can be improved in two ways: 
One way, as pointed out by the authors of \cite{mda}, is to 
use an ECC with better FER-performance. A second way, as we have pointed out here, is to base the implementation on a method of rate-lowering 
resulting in a better
FER-performance (while minimizing the extra classical leakage, see Eq.~\eqref{extraLeakage}). 

In order to isolate and quantitatively demonstrate the second effect, we have studied the performance of MDA-implementations based on the RL-LDPC code \cite{ecc}, where we employed as the method of rate-lowering either bit-revelation ($\text{MDA}_a$) as in \cite{mda} or the method inherent to RL-LDPC codes ($\text{MDA}_b$). 
In different application scenarios, we have shown significant gains of $\text{MDA}_b$ compared to $\text{MDA}_a$ (see Fig.s \ref{fig:plotEx1},\ref{fig:plotEx2} and Eq.s \eqref{iABex1},\eqref{iABex2}, respectively). 
By resolving the performance of the various schemes in terms of $\overline{l}$, the average number of decoding iterations (in addition to the heuristic measure used in \cite{mda}, $\overline{D}$ (see Eq.~\eqref{DBar})), we have shown that the above gains come at no extra cost in decoding complexity (see Fig.s \ref{fig:complexitySKF} and \ref{fig:complexityUsedSKFZoom}).
MDA-gains (in either implementation) compared to the best possible SDA key rate can require a slight cost of 
$\overline{l}$, which can however be eliminated by implementing LLR-inheritance (see Fig.s \ref{fig:complexitySKF} and \ref{fig:complexityUsedSKFZoom}).

	\section*{Acknowledgments}
We would like to thank Kadir Gümüş for correspondence on the work \cite{mda}.	

This research was conducted within the scope of the
projects QuNET and QuNET+ProQuake, funded by the German Federal Ministry
of Research, Technology and Space (BMFTR) in the context of the
federal government’s research framework in IT-security
”Digital. Secure. Sovereign.”.

\section*{Data Availability Statement}
The data that support the findings of this article will be made openly available in \cite{zeno} upon journal submission.

\section*{Appendix}\label{sec:appendix}

\begin{figure*}
	\begin{minipage}{0.98\columnwidth}
		\centering
		\includegraphics[width=\textwidth]{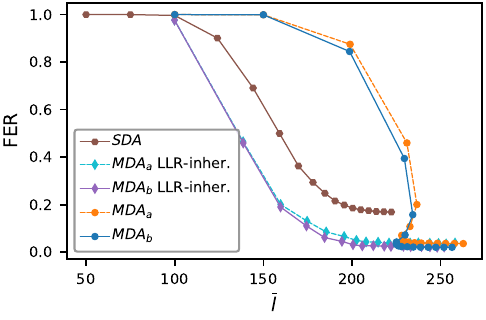}
	\end{minipage}
	\hfill
	\begin{minipage}{0.98\columnwidth}			\centering
		\includegraphics[width=\textwidth]{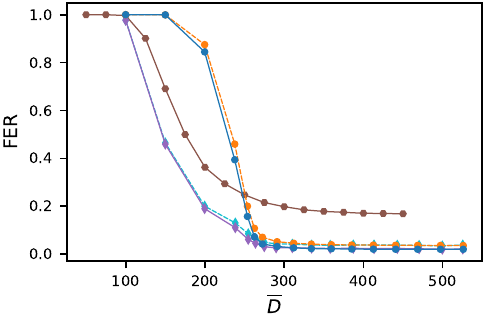}
	\end{minipage}
	\caption{Performance and complexity: FER achieved by the various schemes versus the average number of decoding iterations used, $\overline{l}$, and the average allowed number of decoding iterations, $\overline{D}$ (see Eq.~\eqref{DBar}), respectively. The plotted points result from a variation of $l_{\textit{max}}$ from $50$ to $450$ in steps of $25$.}
	\label{fig:complexityFER}
\end{figure*}

For the simulations in section \ref{sec:simulated}, 
the normalizations are chosen such that, for each quadrature, Bob receives
\begin{equation}
	y = x + z \,, \quad z \sim \frac{\sigma}{\sqrt{2}} \mathcal{N}\left( 0, 1 \right) \,,
\end{equation}
where Alice chooses $x \in \left\{ -1/\sqrt{2}, \,1/\sqrt{2} \right\}$ with equal probability, and $\mathcal{N}\left( 0, 1 \right)$ is standard normal noise. 
Any transmission loss here is effectively taken into account by $\sigma$.
We then employ reverse reconciliation where 
the raw key is given by the signs of Bob's measurements, $\text{sgn}(y)$, and the absolute values of his measurements, $|y|$, constitute side-information in addition to the syndrome of the raw key. 
The frame length required by the ECC \cite{ecc} is given in terms of the code rate as $n = 2\cdot10^4 \, \lfloor r^{-1} \rfloor$.

We use Sum-Product-Algorithm decoding. We acknowledge the use of the decoder implementation from \cite{toolkit} (version 1.0).	
For the LLR-inheritance as well as the early termination of the decoder investigated in \ref{ssec:complexity}, minor modifications of the source code of \cite{toolkit} are necessary.

The simulations in section \ref{sec:simulated} have been carried out with $N_f =2400$ binary frames simulated per plotted point (corresponding to $1200$ QPSK-symbol frames transmitted). The exception are the simulations of Fig.s 
\ref{fig:complexitySKF}
and 
\ref{fig:complexityFER} with $N_f = 9600$ for the applications of MDA that include LLR-inheritance and $N_f = 4800$ for the remaining data in those plots.

As indicated, following  \cite{mda}, for all key rate calculations in this work we calculate the Holevo information based on Eq.~(9) from \cite{zhangHolevo}. Likewise following \cite{mda}, we choose $\eta=0.5$, $\xi=0.01$, $\nu_{el}=0.1$.
The SNR is then given as \cite{zhangHolevo}
\begin{equation}
	\text{SNR} = \frac{1}{\sigma^2} = \frac{T \, V_A}{T \,\xi + 2\frac{1+\nu_{el}}{\eta}} \,.
\end{equation}
For the variation of the SNR in Fig.~\ref{fig:range},
we fix $V_A$ and solely vary the transmission loss $T$, following \cite{mda}.The specific choices of $V_A$ considered in Sec.~\ref{sec:simulated} are motivated as follows. 

The choice $V_A=0.5$ follows that of \cite{mda}. There, in combination with the ECC used there, it leads to $\text{FER}_* \approx 19 \%$. However, the better performance of the ECC used here results in $\text{FER}_* \approx 2 \%$. Thus, we consider $V_A=0.5$ in \ref{sssec:ex2} as a representative case of the scenario $\text{FER}_* \ll 1$.

In order to have a use case comparable with \cite{mda} (i.e., a value of $\text{FER}_*$ comparable to $19 \%$), according to the FER-results of Fig.~\ref{fig:range}, we need to choose a value of $V_A$ selecting $\beta_*=95.5 \%$ corresponding to $\text{FER}_* \approx 16.7 \%$. Among the $V_A$-values achieving this, we then choose the one maximizing $K_*$. With a resolution of $10^{-2}$, this results in $V_A=0.80$. Correspondingly, we consider this choice in \ref{sssec:ex1} as our example for the scenario of a high $\text{FER}_*$.

\end{document}